\documentclass[usenatbib]{mn2e}
\usepackage{epsfig}
\usepackage{amsmath}
\title[The environmental dependence of star formation]{The 2dF Galaxy Redshift Survey: The environmental dependence of galaxy star formation rates near clusters}
\author[Lewis \etal]{
Ian Lewis$^{1,2}$, Michael Balogh$^{3}$, Roberto De Propris$^{4}$, Warrick Couch$^{4}$, Richard Bower$^{3}$, \newauthor
Alison Offer$^{2}$, Joss Bland-Hawthorn$^2$, Ivan K. Baldry$^{5}$, Carlton Baugh$^{3}$, Terry Bridges$^2$,\newauthor
 Russell Cannon$^2$, Shaun Cole$^3$, Matthew Colless$^6$, Chris Collins$^7$, Nicholas Cross$^{6,8}$, \newauthor
Gavin Dalton$^1$, Simon P. Driver$^{6,8}$, George Efstathiou$^9$, Richard S. Ellis$^{10}$,  \newauthor
Carlos S. Frenk$^3$, Karl Glazebrook$^2$, Edward Hawkins$^{11}$, Carole Jackson$^{6}$, \newauthor
Ofer Lahav$^9$, Stuart Lumsden$^{12}$, Steve Maddox$^{11}$, Darren Madgwick$^{9}$, Peder Norberg$^3$, \newauthor
John A. Peacock$^{13}$, Will Percival$^{13}$, Bruce A. Peterson$^6$, Will Sutherland$^{13}$, \newauthor
Keith Taylor$^{10}$\\
$^{1}$Astrophysics, Nuclear and Astrophysics Laboratory, Keble Road, Oxford OX1 3RH, UK\\
$^{2}$Anglo-Australian Observatory, P.O. Box 296, Epping, NSW 1710, Australia \\
$^{3}$Department of Physics, University of Durham, South Road, Durham DH1 3LE, UK\\
$^{4}$School of Physics, University of New South Wales, Sydney 2052, Australia\\
$^{5}$Department of Physics and Astronomy, Johns Hopkins University, 3400 North Charles Street, Baltimore, MD 21218-2686 USA\\
$^6$Research School of Astronomy \& Astrophysics, The Australian National University, Weston Creek, ACT 2611, Australia \\
$^7$Astrophysics Research Institute, Liverpool John Moores University, Twelve 
Quays House, Egerton Wharf, Birkenhead, L14 1LD, UK \\
$^{8}$School of Physics and Astronomy, University of St. Andrews, North Haugh, St Andrews, Fife KY16 9SS, UK\\
$^9$Institute of Astronomy, University of Cambridge, Madingley Road, Cambridge\\ 
$^{10}$California Institute of Technology, Pasadena, CA, 91125-2400, U.S.A.\\
$^{11}$School of Physics and Astronomy, University of Nottingham, University 
Park, Nottingham, NG7 2RD, UK \\
$^{12}$Department of Physics \& Astronomy, E C Stoner Building, Leeds LS2 9JT, UK \\
$^{13}$Institute of Astronomy, University of Edinburgh, Royal Observatory, 
Edinburgh EH9 3HJ, UK \\
}
\date{\today}
\def\2df{2dFGRS}
\def\etal{{ et al.\thinspace}}
\def\gtrsim{\mathrel{\raise0.35ex\hbox{$\scriptstyle >$}\kern-0.6em
\lower0.40ex\hbox{{$\scriptstyle \sim$}}}}
\def\lesssim{\mathrel{\raise0.35ex\hbox{$\scriptstyle <$}\kern-0.6em
\lower0.40ex\hbox{{$\scriptstyle \sim$}}}}

\def\kmsmpc{{\,\rm km\,s^{-1}Mpc^{-1}}}
\begin{document} 
\maketitle
\begin{abstract}
We have measured the equivalent width of the H$\alpha$ emission line for 11006
galaxies brighter than $M_b=-19$ ($\Omega_\Lambda=0.7$, $\Omega_m=0.3$, $H_0=70\kmsmpc$)
at $0.05<z<0.1$ in the 2dF Galaxy Redshift Survey (\2df), in the fields of
seventeen known galaxy clusters.  The limited redshift range ensures that
our results are insensitive to aperture bias, and to residuals from night
sky emission lines.  We use these measurements to trace $\mu^\ast$, the star formation rate 
normalized to $L^\ast$, as a function of distance from the cluster centre, and local projected 
galaxy density.  We find that the distribution of $\mu^\ast$ steadily skews toward larger
values with increasing
distance from the cluster centre, converging to the field distribution 
at distances greater
than $\sim 3$ times the virial radius.  
A correlation
between star formation rate and local projected density is also found,
which is independent of cluster velocity dispersion and disappears at projected densities below $\sim 1$ galaxy (brighter than $M_b=-19$)
per Mpc$^{2}$.  This characteristic scale corresponds approximately to the mean density  
at the cluster virial radius.  The same correlation holds for galaxies more than two virial
radii from the cluster centre.  We conclude that environmental
influences on galaxy properties are not restricted to cluster cores, but are effective
in all groups where the density exceeds this critical value.  The present day abundance of such
systems, and the strong evolution of this abundance, makes it likely that hierarchical growth
of structure plays a significant role in decreasing the global average star formation rate.
Finally, the low star formation rates well beyond the virialised cluster rule out severe physical
processes, such as ram pressure stripping of disk gas, as being completely responsible for the variations
in galaxy properties with environment.  
\end{abstract}
\section{Introduction}
The effect of local environment on galaxy evolution in general is not well understood.
Studies of environmental effects in the past have been largely devoted to the study of galaxies
in the cores of rich clusters, which differ so dramatically from more common
galaxies \citep[e.g. ][]{Dressler,DTS,CS87,B+97,PSG,P+99,MW00,C+01,Solanes01}.  
However, galaxies in cluster cores comprise only a small fraction of the
stellar content within the universe, and thus it is not obvious that
the processes which effect these galaxies are important for galaxy evolution
in general.

More recently, however, work has begun to show that star formation is suppressed in cluster
galaxies far from the core.  From the CNOC1 cluster sample, \citet{B+97,B+98} found that the mean cluster galaxy
star formation rate may be suppressed as far as  twice the virial radius ($R_v$) from the cluster centre,
relative to a field sample selected in the foreground and background of the clusters.
However, the data at large radii were sparse, and being derived from only a few clusters
were sensitive to the effects of substructure and non-sphericity.  Thus it is not possible
to draw strong conclusions about the
relative cluster galaxy star formation rate beyond the $R_v$ from these data.  Wide field photometric
analysis of clusters using Subaru has recently suggested that the tight red sequence of early type
galaxies first presents itself in small groups of galaxies within the infall region of
the massive cluster Cl0939+47 at $z=0.39$ \citep{Kodama_cl0939}.   This is the first work
to suggest that a ``critical'' environment for galaxy evolution exists.
A larger survey, designed
specifically to study the outer regions of clusters is the
Las Campanas/Anglo Australian Observatory Rich Cluster Survey (LARCS), a 
sample of 17 rich, X-ray bright clusters, with photometry and spectroscopy extending
out to very large radii ($\sim 6$ Mpc).  Early results confirm the radial
gradient in photometric and spectroscopic properties out to the virial radius and, perhaps,
beyond \citep{OHely,kap1,kap2}.  

It therefore seems likely that galaxy star formation rates are reduced
before they are accreted by a cluster, for example in smaller groups.  
If this is the case, the implications could be profound, as  
most galaxies at the present day are in groups \citep{TG72,Cfa3,Tully87,CNOC_groups}; if environmental processes
are important in these regions, they will clearly be reflected in the evolution of the
universe as a whole.  As structure builds up in the universe, more and more
galaxies can be found in groups and, if these environments serve to terminate
star formation, the mean star formation rate of the universe will decline.
This might explain at least part of the observed decline in global star formation
with cosmic time \citep{L96,Madau,Cowie+99}.

The 2dF galaxy redshift survey (\2df) allows the unprecedented opportunity to study
the spectroscopic properties of galaxies at an arbitrarily large distance from any
given cluster.  The details of the survey strategy are given elsewhere \citep{2dF_colless},
but summarized briefly in Section~\ref{sec-specdat}.
Analysis of the whole sample will allow a definitive study of any correlation between
spectral properties (i.e. emission line strength) as a function of a continuous
variable like local density.  For this preliminary study, we are specifically interested
in establishing precisely where galaxies in the vicinity of known clusters begin to
exhibit properties which differ from those of the average galaxy.   We base this on a sample
of 17 known rich clusters within the \2df, from the catalogue of \citet{deP_clus}.

Our cluster selection, galaxy sample, and star formation rate measurements are described
in Section~\ref{sec-data}.  In Section~\ref{sec-results} we show the trend of increasing star formation
activity with both increasing cluster-centric distance, and decreasing local projected
density.  This is compared with numerical models in Section~\ref{sec-discuss}.  We summarize
our findings in Section~\ref{sec-conc}.  Throughout this paper, we
use a cosmology with $\Omega_\Lambda=0.7$, $\Omega_m=0.3$, $H_0=70\kmsmpc$.
We use the symbol $M_b$ to denote absolute magnitudes
measured in the 2dFGRS photographic blue system.

\section {Data Analysis}\label{sec-data}
\subsection{Spectroscopic Data}\label{sec-specdat}
The 2dF Galaxy Redshift Survey has obtained over 220\,000 spectra of 
galaxies located in two contiguous declination strips, plus 99
randomly located fields. One strip is in the southern Galactic 
hemisphere and covers approximately 80${^\circ}\times15{^\circ}$ centred 
close to the SGP. The other strip is in the northern Galactic hemisphere 
and covers 75${^\circ}\times10{^\circ}$. The 99 random fields are located 
over the entire region of the APM galaxy catalogue in the southern 
Galactic hemisphere outside of the main survey strip. Full details of the 
survey strategy are given in \citet{2dF_colless}.

The survey spectra cover the wavelength range 3600--8000\AA\ at 9\AA\ 
resolution. Only the wavelength range of 3600--7700\AA\ is used during the 
line fitting procedure due to poor signal to noise and strong sky emission 
in the red part of the spectrum. The wide wavelength range is made 
possible by the use of an atmospheric dispersion compensator (ADC) within 
the 2dF instrument \citep{2dF}.

\subsection{Cluster Selection}
We select 17 clusters from the catalogue of \citet{deP_clus}, in which clusters
from the Abell catalogues \citep{A58,ACO}, the APM \citep{APM} and the EDCC \citep{EDCC}
were cross-referenced with the \2df.  
This catalogue is still partially incomplete, but the completeness is generally
better than 75\% within $\sim 5$ Mpc of the cluster centres.  The mean redshift and velocity dispersions of the clusters
in this catalogue have been recomputed from the \2df\ spectra, and the
cluster centroid is taken to be the brightest cluster galaxy
with early-type morphology, identified from POSS plates.

\begin{figure*}
\leavevmode \epsfysize=8cm\epsfbox{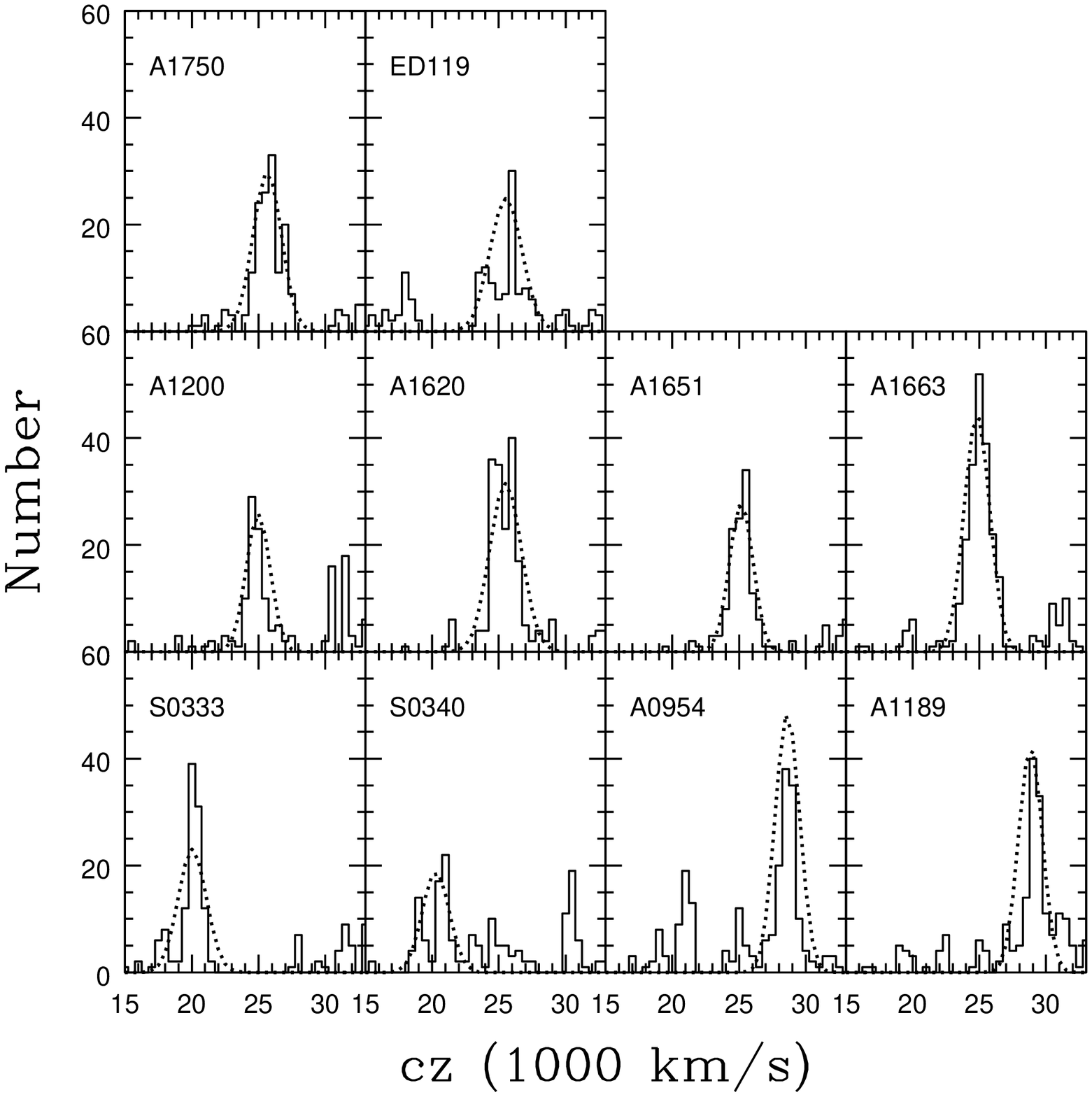}\epsfysize=8cm\epsfbox{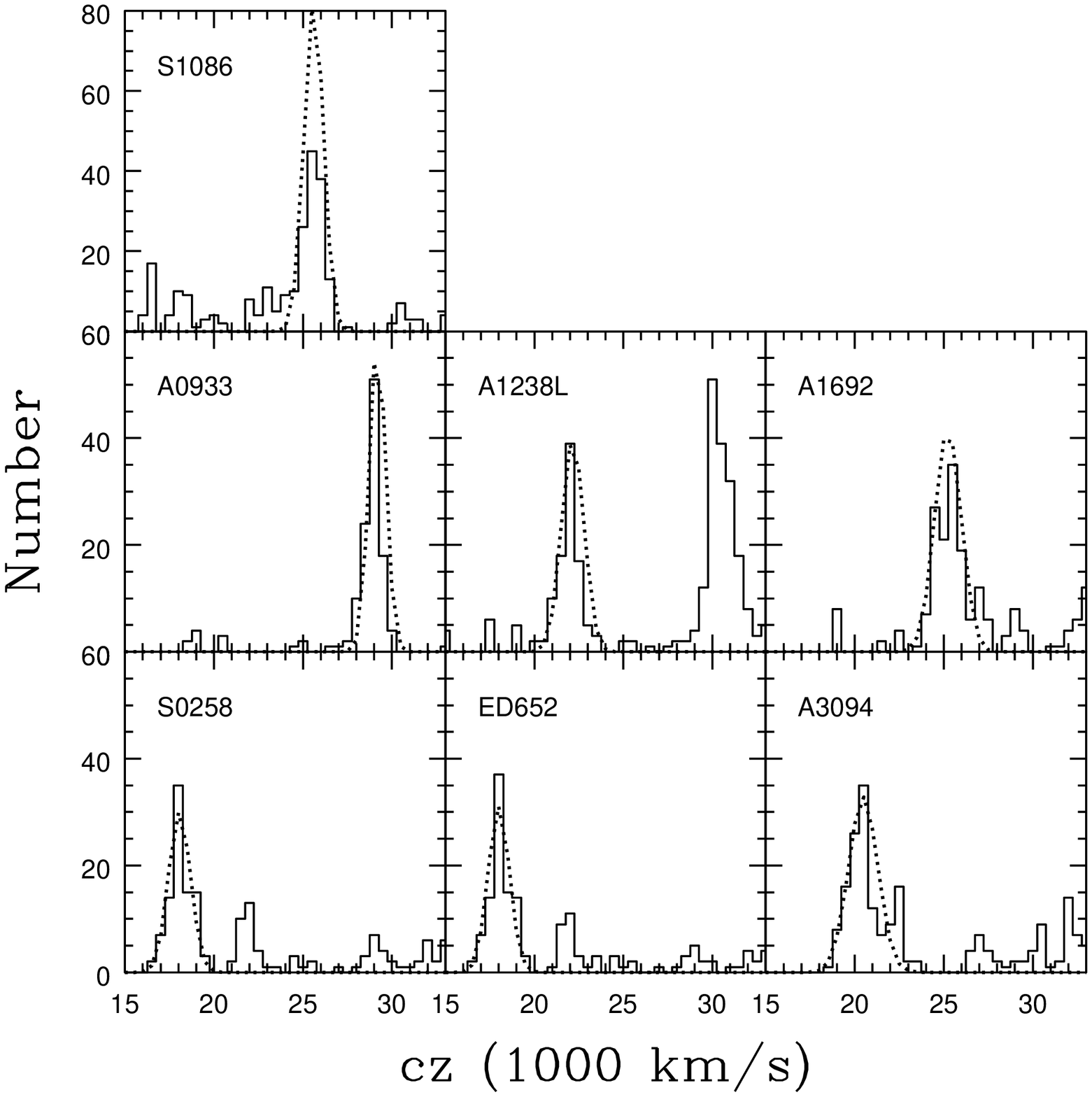}
\caption{Redshift histograms for the seventeen clusters used in this study.  Only galaxies
brighter than $M_b=-19$ and within 5 Mpc of the cluster centre are included.
The {\it dotted} line overlayed on each histogram represents a Gaussian with
a central redshift ($cz$) and velocity dispersion as tabulated in Table \ref{tab-clus}.
{\bf Left: }The ten clusters with velocity dispersion $\sigma>800$~km~s$^{-1}$.  {\bf Right: }The remaining seven clusters with
$\sigma<800$~km~s$^{-1}$.
\label{fig-himass}}
\end{figure*}

For this analysis, we extract from the \2df\ all galaxies within $\sim20$~Mpc
of the centre of 17 clusters, selected to lie at $18\,000\;{\rm km\,s}^{-1}<cz<29\,000\;{\rm km\,s}^{-1}$.
The lower
velocity bound  is chosen to limit the angular size to a reasonably small,
manageable value; the upper limit is defined as the velocity at which H$\alpha$ is 
redshifted into the first set of strong night-sky OH emission lines.  Ten clusters
were selected to have velocity dispersions $\sigma>800$ km/s, while the
remaining seven are systems with $400$ km s$^{-1}$$<\sigma<800$ km s$^{-1}$.
The redshift histograms for the 17 clusters, including all galaxies
brighter than $M_b=-19$ within 5 Mpc (projected) of the centre, are shown in Fig.~\ref{fig-himass}.
Details of the clusters, including their redshifts ($cz$), velocity dispersions
($\sigma$), number of cluster members brighter than $M_b=-19$, 
and completeness (within 5 Mpc), are summarized in Table~\ref{tab-clus}.
\citet{deP_clus} resolved Abell 1238 into two clusters aligned along the
line of sight; we here consider the lower redshift cluster, designated Abell 1238L.
The cluster centres and velocity dispersions are generally better
determined than they appear in Fig.~\ref{fig-himass}, as they are computed including
fainter galaxies over a smaller projected area (where the contrast with the field is
greater).  

%
%
\begin{table*} 
\caption{ {\sc \label{tab-clus} The seventeen clusters used in this study}} 
\vspace{0.1cm}
{
\begin{tabular}{lcccccccc} 
\hline\hline
\noalign{\smallskip}
Name      &   R.A.\   & Dec.\       & $cz$ & $N_{\rm mem}$&$\sigma$& Completeness& $R_v$ & $R_v$ (alt.)\cr 
          & \multispan2{\hfil (B1950)\hfil }   &  (km s$^{-1}$)& &(km s$^{-1}$)&(within 5 Mpc)&(Mpc)&(Mpc)\cr 
\noalign{\smallskip}
\noalign{\hrule}
S0258&02:23:33.21&$-$29:50:26.9&18060&31&583&0.72&1.6&1.9\cr
ED652&02:25:11.88&$-$29:51:00.7&18001&21&564&0.75&1.4&1.8\cr
A3094&03:09:16.42&$-$27:07:08.4&20475&63 &774&0.84&2.0&2.4\cr
S0333&03:13:04.34&$-$29:25:41.3&20042&40 &998&0.90&1.6&3.2\cr
S0340&03:17:55.68&$-$27:11:45.6&20281&18 &939&0.87&1.2&3.0\cr
A0933&10:05:14.50&+00:45:25.7&29180&72 &420&0.54&2.4&1.3\cr
A0954&10:11:11.10&+00:07:40.2&28622&74 &832&0.77&2.2&2.5\cr
A1189&11:08:30.14&+01:21:42.6&28824&51 &814&0.77&1.9&2.5\cr
A1200&11:10:03.25&$-$02:56:27.6&24970&38 &825&0.83&1.7&2.6\cr
A1238L&11:20:20.36&+01:23:19.4&22160&53 &586&0.82&1.9&1.8\cr
A1620&12:47:29.78&$-$01:16:07.1&25513&51 &1095&0.89&1.8&3.4\cr
A1651&12:56:47.48&$-$03:55:36.9&25152&46 &817&0.47&2.1&2.5\cr
A1663&13:00:18.05&$-$02:14:57.7&24827&75 &884&0.80&2.1&2.7\cr
A1692&13:09:41.25&$-$00:39:59.7&25235&49 &686&0.80&1.8&2.1\cr
A1750&13:28:36.52&$-$01:28:15.9&25647&83 &981&0.62&2.4&3.0\cr
ED119&22:13:32.57&$-$25:55:10.7&25546&38 &1112&0.84&1.7&3.4\cr
S1086&23:02:06.51&$-$32:49:14.8&25605&74 &502&0.53&2.4&1.5\cr
\noalign{\hrule}
\end{tabular}
}
\end{table*}

\subsection{H$\alpha$ measurements}\label{sec-measure}
All of the measurements of equivalent width have been performed using a
completely automatic procedure. For each spectrum we remove the continuum
by subtracting the median over a 133\AA\ (31 pixel) wide window
after first excluding known absorption and emission
line regions by making use of the known galaxy redshift. Bad pixels and
sky line residuals and the atmospheric and fibre absorption bands are
also excluded from the continuum fitting.

Both emission and absorption lines are fitted with Gaussian profiles which are
adequate for most of the emission lines and cores of the absorption
lines.  
Up to 20 individual absorption and emission lines are fitted
simultaneously using a modified Levenberg-Marquardt algorithm. The width and
height of each line are fitted
together with a small perturbation of the observed redshift. Some
lines were constrained to be emission or absorption. Others were
allowed to be either. Note that the
relative wavelength spacing of all lines remains fixed, but the fitted
redshift is allowed to vary slightly (typically $\Delta z \sim 0.00025$, and always
$\Delta z < 0.005$). By fitting many lines simultaneously we
avoid individual line fits shifting to the nearest available peak or dip
in the spectrum. By fitting both absorption and emission lines we ensure
that the method is robust to the redshift solution whatever type of
spectrum is being fitted.

\begin{figure*}
\psfig{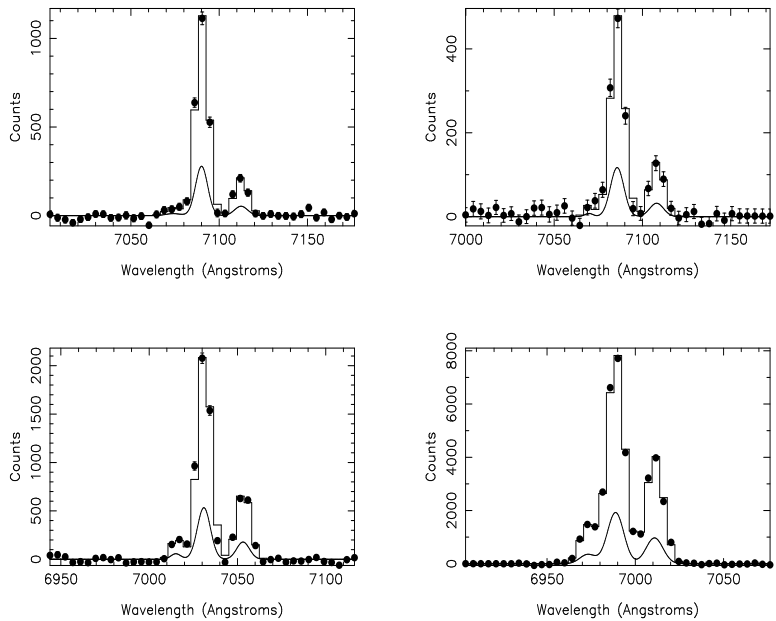}
\caption{Four examples of the Gaussian line-fits to the spectra, for varying
strengths of [N{\sc ii}] and signal-to-noise ratio.  Each plot shows the observed 
low resolution data as a series of filled circles with rms error bars. The 
modelled Gaussian fit is shown as a smooth curve and 
the modelled Gaussian sampled at the same resolution as the observed data 
is shown as a histogram.
\label{fig-ijl}}
\end{figure*}

With this technique of simultaneous line fitting it is possible to allow
for line blends by simply requesting two or more lines to be fitted to
the blend.  For example H$\beta$
is best fitted by a combination of a narrow emission and a broad absorption
line, and the H$\alpha$ emission line can be accurately deblended from the adjacent
[N{\sc ii}]$\lambda6548$\AA\ and [N{\sc ii}]$\lambda$6583\AA\ lines, 
despite the 9\AA\ resolution of the spectra.  The [N{\sc ii}] lines are
constrained to be in emission while the H$\alpha$ line may be either
emission or absorption.  To fit the Gaussian profile to the data points a consideration has to be 
made for the effect of the undersampling of the data. The solution is to 
model a Gaussian profile which, when undersampled, fits the observed data 
closely.  Fig.~\ref{fig-ijl} shows the resulting fit for four spectra with
varying [N{\sc ii}]/H$\alpha$ ratios, and 
demonstrates the effect of the undersampling.

After line fitting, the parameters of the fit (amplitude, sigma and area)
and the rms residuals are used to
classify the quality of the line fit.  Usual reasons for rejecting a fit are
if the line is too narrow (e.g. a noise spike or residual cosmic ray hit),
or too broad (for a forbidden line).  Some combinations of lines are also rejected, for
example H$\alpha$ absorption combined with [N{\sc ii}] emission. 
Lines which are too weak for a good fit
are also flagged; however they are not rejected from the analysis, so that
we retain a dispersion in the flux which reflects the measurement uncertainties.  
Partial failures are flagged often due to large rms
residuals to the fit (broad non-thermal emission lines are poorly fitted
by Gaussian profiles) or a poor wavelength calibration at the blue end
of the spectrum, which can lead to a poor line profile for [O{\sc ii}]. The latter
is usually the case when the observed spectrum was close to the edge of
the CCD. Care is also taken when a bad pixel has been masked out from
the spectrum within 2$\sigma$ of the line centroid.

Equivalent widths are then simply calculated using the continuum fit and the
measured line flux. A small number of spectra are degraded
by poor sky subtraction at the data reduction stage, which can result
in a negative continuum, making the EW meaningless. These cases
can be easily removed from subsequent analysis.

\subsection{Sample Selection and Star formation Rates}\label{sec-sample}
\2df\ spectra within 20 Mpc of each cluster centre were extracted from the database.
The extreme ends of spectra taken earlier than August 1999 are severely affected by 
problems with the ADC \citep{2dF}; hence we restrict our analysis to data
taken after this date.  This leaves us with 53018 galaxies, but to limit the effect of 
aperture bias and sky-subtraction residuals, we restrict the sample to those galaxies which
lie within $0.05<z<0.1$.  Within this redshift range, the galaxy sample is complete
to $M_b=-19$, and we adopt this as our luminosity limit.
This leaves us with 12020 galaxies.
For computations of star formation rates, we exclude galaxies in which the continuum was negative,
or a Gaussian was a poor fit to the line (see Section~\ref{sec-measure}).  This removes
an additional 734 galaxies from the sample ($\sim 6$\%).  Finally, for galaxies with a significant
H$\alpha$ equivalent width ($W_{\rm H\alpha}>10$ \AA) we exclude galaxies in which the 
equivalent width of the adjacent [N{\sc ii}]$\lambda$6583 line is greater than 0.55$W_{\rm H\alpha}$.
These 280 galaxies (2.3\% of the sample) are likely to have a significant non-thermal 
component \citep{VO87}.  This leaves us with a final sample of 11006 galaxies.
We take cluster members to be those within 3$\sigma$, where $\sigma$ is the cluster
velocity dispersion determined by \citet{deP_clus}, shown in Table~\ref{tab-clus}.   The number of
such members within the virial radius (see below) is denoted $N_{\rm mem}$ in the table.
Note that for the three highest redshift clusters A0933, A0954 and A1189, the highest velocity
members are not included due to our overall redshift cut ($0.05<z<0.1$); however, all galaxies
within 2$\sigma$ are still available.
5829 galaxies in our final sample are thus defined as cluster members.

\citet{K83,K92} derived a conversion from H$\alpha$ luminosity to star formation rate, under
the assumptions of Case B recombination, no escape of Lyman-$\alpha$ photons, and
a Salpeter-like initial mass function.  This may underestimate the current star formation
rate by a small factor, due to extinction in the line-emitting regions \citep{CL}.  Also,
if the nature of star formation is burst-like, the instantaneous star formation rate may
not be representative of the average over even short ($\sim 100$ Myr) timescales \citep{S+01}.
However, neither of these effects are likely to affect a comparison of galaxy populations with
similar luminosity functions, as is the case in the present work.

Since the \2df\ spectra are not flux calibrated, we cannot derive H$\alpha$ luminosities,
or star formation rates.  However, after making a small (2\AA) correction for the underlying
stellar absorption, we can use the equivalent widths to calculate the 
star formation rate normalised to a fiducial luminosity (essentially a star formation rate
per unit normalized luminosity). 
If $\mu$ is the star formation rate in units
of $M_\odot$ yr $^{-1}$ and $L_{\rm H\alpha}$ is the total luminosity of the $H\alpha$ emission line
in ergs~s$^{-1}$, we can define
\begin{equation}
\eta=\mu/L_{\rm H\alpha}.
\end{equation}
We will use the ``average'' conversion factor of 
$\eta=7.9\times10^{-42}$ $M_\odot$ s yr$^{-1}$ ergs$^{-1}$\citep{K92}.
The equivalent width of H$\alpha$, corrected for stellar absorption, is given by
\begin{equation}
W_{\rm H\alpha}\approx L_{\rm H\alpha}/L_c,
\end{equation} 
where $L_c$ is the continuum luminosity in units of ergs~s$^{-1}$\AA$^{-1}$.  We can then calculate
$\mu^\ast$ as
\begin{equation}
\mu^\ast = {\mu\over L_c/L^\ast}=\eta W_{\rm H\alpha} L^\ast, 
\end{equation}
where $L^\ast$ is a characteristic luminosity, for normalisation, in units of ergs~s$^{-1}$\AA$^{-1}$.
We take $L^\ast$ to correspond to the knee in the luminosity function in the $r^\prime$ band (near
rest-frame H$\alpha$), as determined
by \citet{Sloan_lf}, $M_R=-21.8$ ( $\Omega_\Lambda=0.7$, $\Omega_m=0.3$, $h=0.7$), or $L^\ast=1.1\times10^{40}$ ergs~s$^{-1}$\AA$^{-1}$.
Therefore, we have
\begin{equation}
\mu^\ast = 0.087 W_{\rm H\alpha}, 
\end{equation}
which gives the star formation rate, in units of $M_\odot$yr$^{-1}$, normalized
to $L^\ast$.

We measure the projected distance of each galaxy from the cluster, as defined by
the brightest central galaxy.  In some cases the cluster membership of a galaxy is
ambiguous, because it lies within 20 Mpc and the 3$\sigma$ redshift limits of more than one cluster
(e.g. clusters S0258 and ED652).  In this case, the galaxy is
assumed to belong to the cluster which is nearest in projected distance.  

\begin{figure*}
\leavevmode \epsfysize=10cm\epsfbox{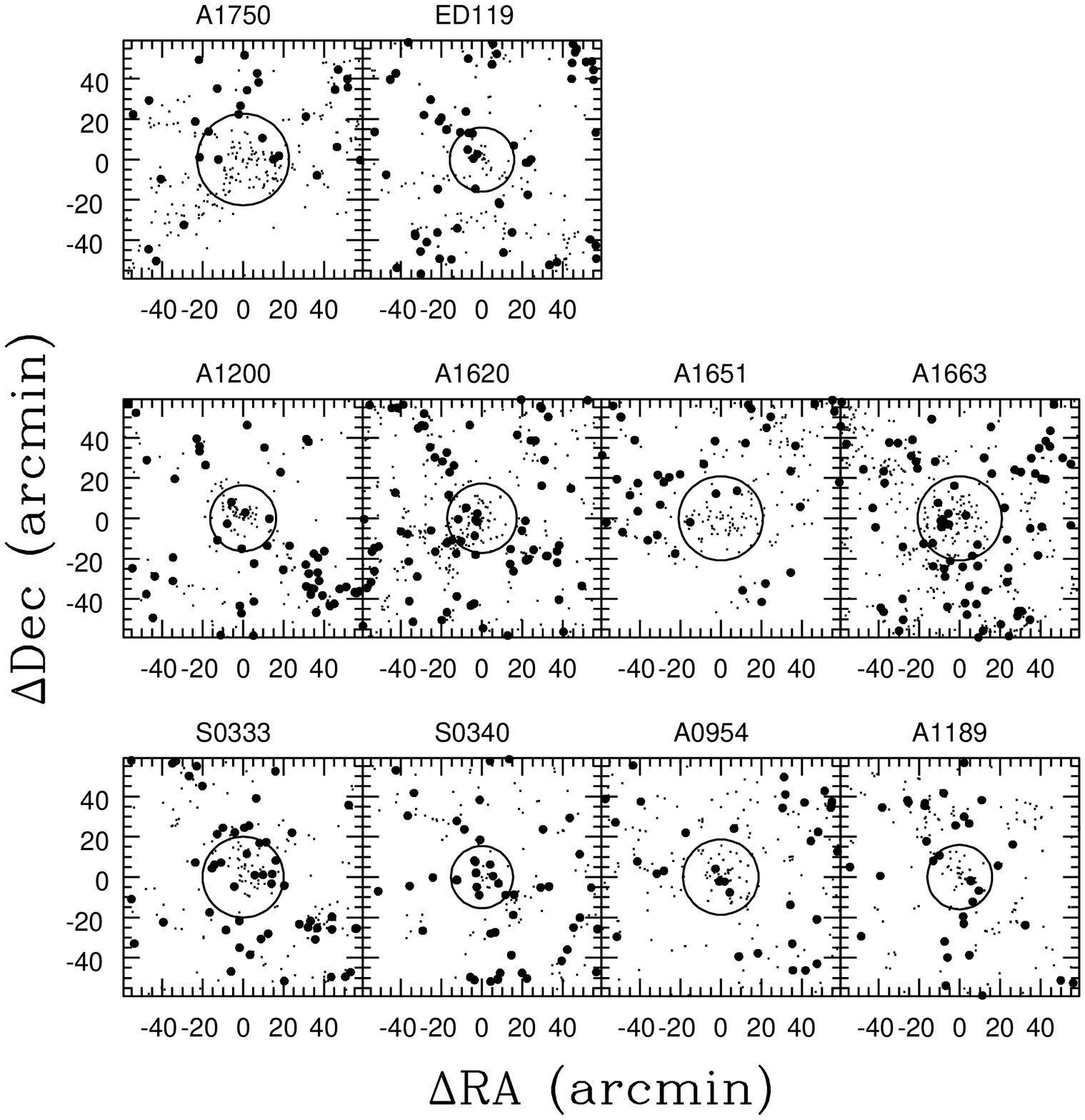}\epsfysize=10cm\epsfbox{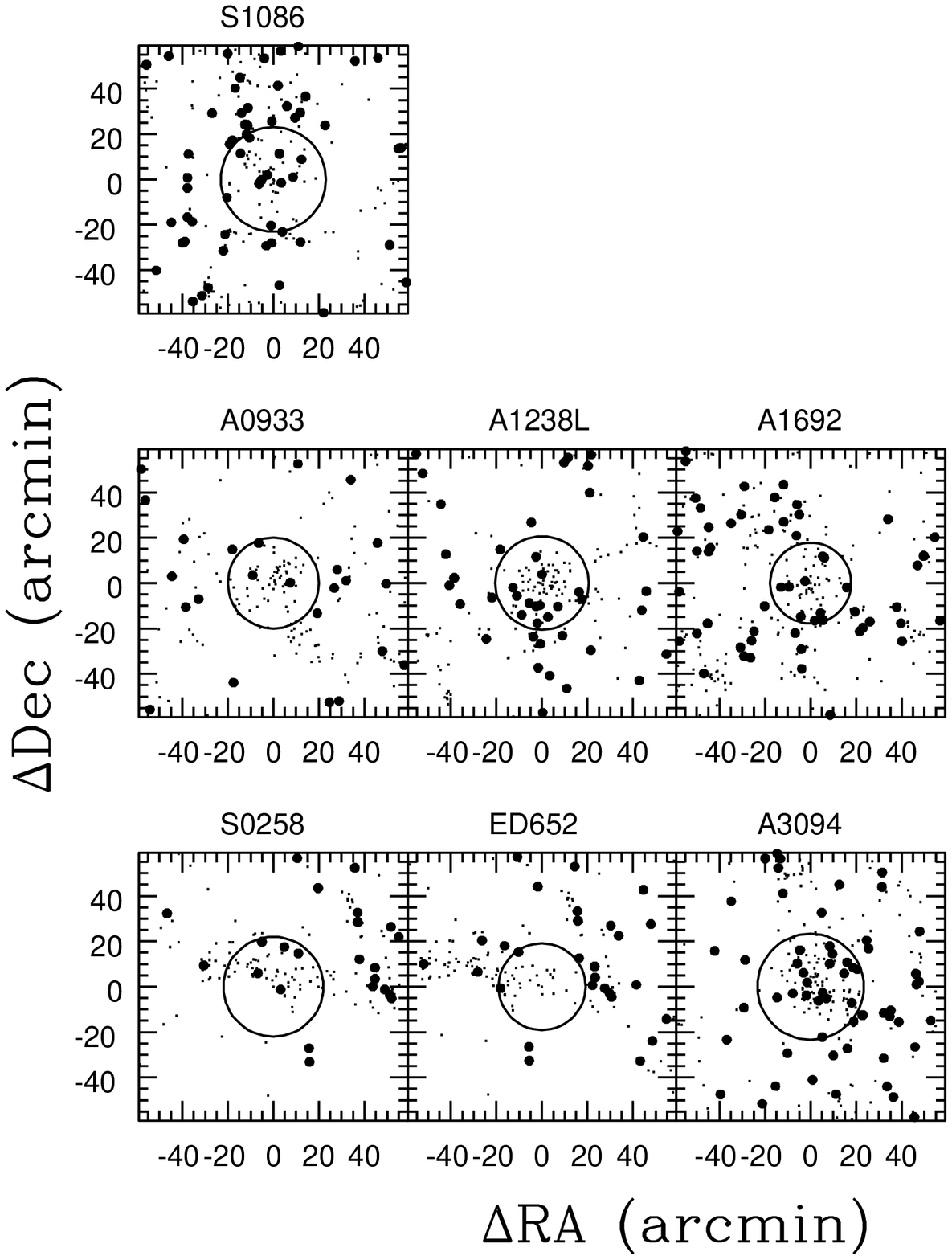}
\caption{Spatial distributions for cluster members within 1 degree of the cluster
centre.  The {\it filled circles} are galaxies with $W_{\rm H\alpha}>20$\AA.  The large
circle in each panel traces the estimated virial radius for the cluster.
{\bf Left: }The ten clusters with velocity dispersion $\sigma>800$~km~s$^{-1}$.  {\bf Right: }The remaining seven clusters with
$\sigma<800$~km~s$^{-1}$.
\label{fig-spatial}}
\end{figure*}

In order to put all the clusters (which span more
than a factor of two in velocity dispersion) on a common scale, and to facilitate
comparison with theory, we need to relate projected distances to the virial radius, $R_v$,
of the cluster. 
We show the spatial distribution of the cluster members within 1 degree of the
centre for each cluster in Fig.~\ref{fig-spatial}.   
From this figure it is evident that many of our clusters are not 
spherically symmetric.
Thus we must be cautious in our interpretation of $R_v$ as a physically meaningful
scale, particularly when considering individual clusters.  

The definition of $R_v$ is 
\begin{equation}\label{eqn-rv}
\bar{\rho}(<R_v)=\Delta_c (z)\rho_c(z)=\Delta_c(z)\rho_b(z)/\Omega_m(z),
\end{equation}
where $\bar{\rho}(<R_v)$ is the mean cluster mass density within $R_v$, $\rho_c$ and
$\rho_b$ are  the critical density and mean background mass density, respectively, and
$\Delta_c$ is the redshift-dependent contrast parameter, determined from spherical
collapse theory.   For
a flat $\Omega_m=1$ universe, $\Delta_c=178$; for our adopted cosmology at $z=0.07$, $\Delta_c\approx 107$
\citep{ECF}, and $\Omega_m(z)=0.343$, so $\Delta_c(z)/\Omega_m(z)=312$.  We will assume that the number density of galaxies
is directly proportional to the dark matter density, independent of scale or galaxy luminosity.  In this
case, we can take the mean background density $\rho_b$ from the luminosity function.
Integrating the best fit Schechter function from \citet{2dF-LFBVD}, we find that the number density of
galaxies brighter than $M_b=-19$ is $\rho_b=0.0076$~Mpc$^{-3}$ ($h=0.7$).  We determine $\bar{\rho}(<R_v)$ by
counting the number of cluster members $N$ within $R_v$ (weighting by the completeness given in Table~\ref{tab-clus})
and assuming a spherical cluster geometry, so
$\bar{\rho}(<R_v)=3N/(4\pi R_v^3)$.  Substituting this into Equation~\ref{eqn-rv}, 
we need to solve $R_v=0.465 N^{1/3}$.  This is done iteratively, by first estimating $R_v$, 
counting the number of members $N$ within $R_v$, and then recomputing $R_v$.  This is repeated
until the solution converges, usually within $\sim 3$ iterations.  These measurements of
$R_v$ are given in column 7 of Table~\ref{tab-clus}.
 
Alternatively, the virial radius can be determined directly from the velocity
dispersion, under various assumptions, as
outlined in \citet{Girardi98}.
If $M_v$ is the virialised mass, and $R_v$ is the cluster virial radius, we have
\begin{equation}
\Delta_c={{3 M_v} \over{4\pi \rho_c R_v^3}}. 
\end{equation}
The virial mass can be related to the velocity dispersion $\sigma$ and $R_v$ under the assumption
of spherical symmetry, through
\begin{equation}
M_v=3G^{-1}\sigma^2R_v,
\end{equation}
so we have
\begin{equation}
R_v={3\sigma \over 4\pi G \rho_c \Delta_c}=\sqrt{6 \over \Delta_c}{\sigma /H_0}.
\end{equation}
For either a flat, $\Omega_m=1$ cosmology (with $h=0.5$) or the $\Omega_\Lambda$-dominated 
cosmology we have adopted ($h=0.7$), the virial radius in Mpc is
$R_v\approx 3.5 \sigma (1+z)^{-1.5}$, for $\sigma$ in units of $1\,000\;{\rm km\,s}^{-1}$.

For 9 of the 17 clusters, this calculation (listed as $R_v$ (alt.) in Table~\ref{tab-clus})
agrees with the previous one to within
$\sim 20$\%.  For most of the remaining cases, where there is a large discrepancy between
the two measurements of $R_v$, the velocity histograms are significantly non-Gaussian,
and thus the velocity dispersion is likely to be a poor tracer of the mass.  For this
reason, we will always adopt the first calculation of $R_v$ as the most likely to be correct.
Moreover, this occaisional discrepancy, and the non-Gaussianity of the corresponding velocity histograms,
likely implies that the computed velocity dispersions are not always simply
related to the virialised mass.  For example, some clusters (ED119, S0333, S0340) may
have velocity dispersions which are artificially inflated by the presence of 
foreground and background structures.
Thus, our division of the sample into two based on velocity dispersion
may not reflect a perfect division into low- and high-mass clusters.

\begin{figure}
\leavevmode \epsfysize=8cm \epsfbox{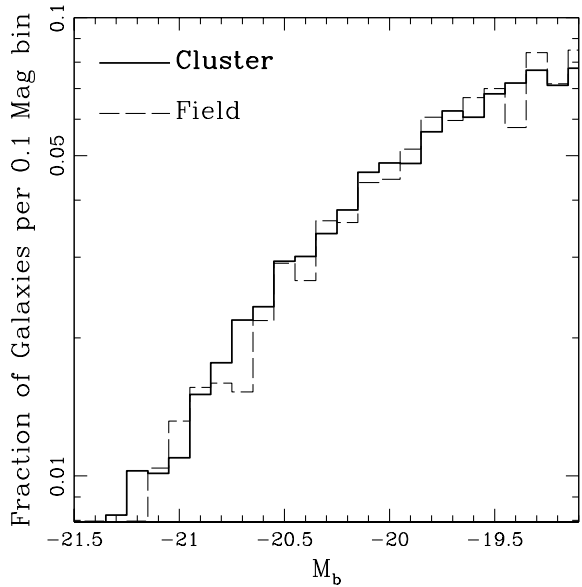}
\caption{Luminosity functions of the cluster sample {\it (solid histogram)} and the
field sample {\it (dashed histogram)}.  
\label{fig-lfunc}}
\end{figure}

We will draw the reference field population from the 2400 galaxies more than
$6\sigma$ from the cluster redshift; i.e. in the foreground and background of the
clusters.  Due to the small redshift range considered, $0.05<z<0.1$, and the use of
an absolute luminosity limit, the field sample is also volume limited.  The
luminosity function of the field sample is comparable to
that of the cluster sample, as shown in Fig.~\ref{fig-lfunc} \citep[see also][]{deP_clus}.

\section{Results}\label{sec-results}
\subsection{General Cluster Properties}
\begin{figure}
\leavevmode \epsfysize=8cm \epsfbox{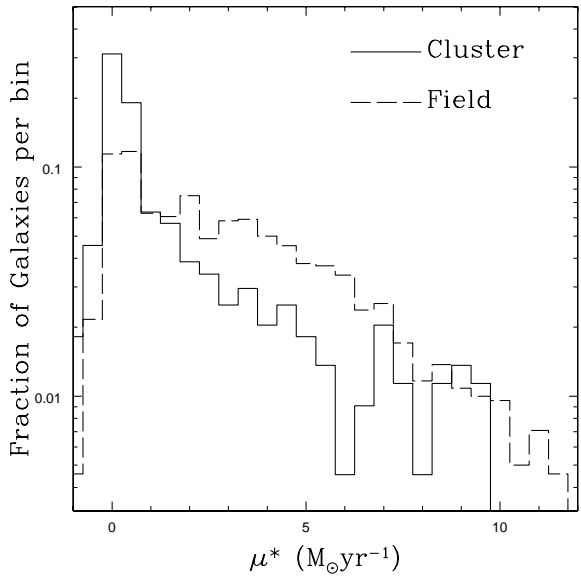}
\caption{The distribution of star formation rate per unit luminosity,
in the cluster and field samples.  The cluster sample is limited to galaxies
within the virial radius.
\label{fig-sfrdist}}
\end{figure}

In Fig.~\ref{fig-sfrdist} we show the distribution of normalized star formation rate, $\mu^\ast$,
in the cluster and field samples, excluding galaxies with relatively strong [N{\sc ii}]$\lambda$6583 emission
(see Section~\ref{sec-sample}).
The cluster sample is limited to the 440 members within $R_v$,
while the field sample is drawn from the 2400 galaxies beyond 6$\sigma$ in velocity.
{\it The difference between the distributions is highly significant\footnote{The probability that
the two distributions are not drawn from the same population is $>99.999$\%  as determined by a
Kolmogorov-Smirnov test.}, with the 
field galaxy population weighted toward galaxies with
stronger star formation}.

\subsection{Radial Dependences}
\begin{figure*}
\leavevmode \epsfysize=10cm \epsfbox{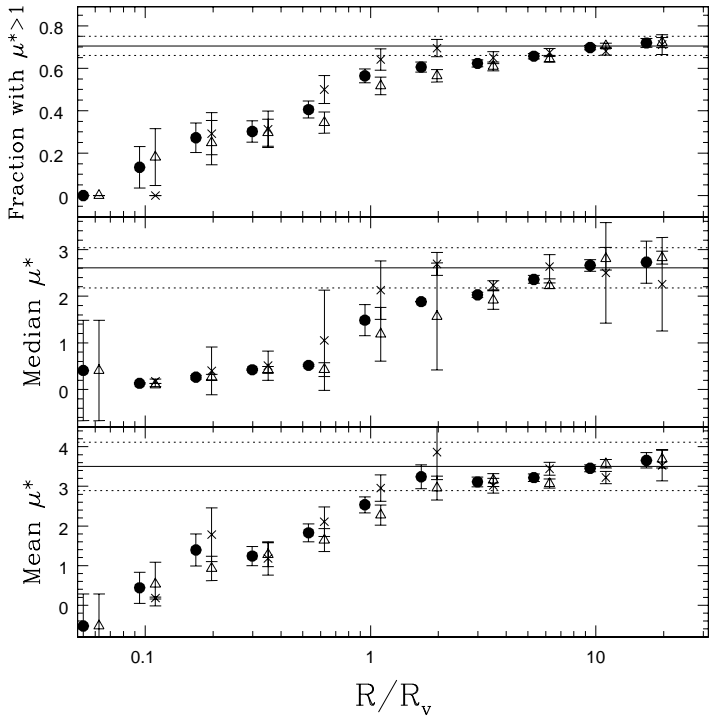}
\caption{The mean ({\it bottom panel}) and median ({\it middle panel}) value of
$\mu^\ast$ in the cluster sample, as a function of radius.   
In the {\it top panel} we show the fraction of galaxies
with $\mu^\ast>1 M_\odot$ yr$^{-1}$.  Error bars are a jackknife resampling estimate.
{\it Solid points} represent the full galaxy sample, while the 
{\it triangles} and  {\it crosses} represent only the clusters
with $\sigma$ greater than or less than 800 km s$^{-1}$, respectively (offset
for clarity).
Only points in which the radial bin contains at least three galaxies are shown.
The horizontal, {\it solid line} represents the value of each statistic in the
field sample.  The {\it dotted lines} which bracket the line are an estimate
of the 1$-\sigma$ field to field standard deviation, for independent samples
of 200 galaxies.
\label{fig-sfrrad}}
\end{figure*}

It is well known that star formation activity in clusters increases with
distance from the centre \citep{B+97,B+98,PSG}.  
In Fig.~\ref{fig-sfrrad} we show how the mean and median value of $\mu^\ast$
depend on radius in our cluster sample, and compare that with the
field value.  We also show, in the third panel, the fraction of galaxies with
$\mu^\ast>1$, which represent the tail of the distribution, comprised of galaxies that are
currently forming stars at a high rate relative to their luminosity.  
The sample is also broken up into clusters with high ($\sigma>800$ km s$^{-1}$, triangles)
and low velocity dispersions ($\sigma<800$ km s$^{-1}$, crosses).  The properties of
the field sample are shown as the horizontal, solid line.  The dashed lines bracketing
the field line represent the 1-$\sigma$ standard deviation from field to field,
computed by ordering the field galaxies in right ascension and treating every 200
galaxies as an individual sample.  This gives some estimate of the expected cosmological
variance in the field value.

All three statistics demonstrate that the cluster distribution of $\mu^\ast$ becomes
equivalent to the field value only well outside the virial radius,
at $R\gtrsim 3 R_v$, in excellent agreement with preliminary results from
the Sloan Digital Sky Survey \citep{Sloan_sfr}.  The implications of this are that
a representative sample of field galaxies cannot be obtained within $\lesssim 6$ Mpc
of the cluster core.  Thus, photometric studies of clusters which attempt a statistical
background subtraction by taking the field from the cluster outskirts \citep[e.g.][]{KB01,kap2} 
are not subtracting enough star-forming galaxies, and artificially inflating the number
of blue galaxies within the cluster.

Since many of the clusters are not spherically symmetric, the interpretation 
of radial gradients, and the physical meaning of $R_v$, is not straightforward.
From Fig.~\ref{fig-spatial} it is clear that there is often considerable structure, both
within and without the virial radius.  Furthermore, the galaxies with strongest H$\alpha$
emission (solid points in Fig.~\ref{fig-spatial}) appear to be spread evenly
throughout the field, avoiding the densest regions, regardless of clustercentric distance.
Thus, in the following section we consider the correlation between star formation rate
and local density.



\subsection{Density Dependences}\label{sec-density}
There has been controversy over whether or not galaxy populations correlate
most closely with cluster-centric radius \citep{WGJ} or local density
\citep{Dressler,PG84}.  If radius is the primary determinant anywhere, it is
most likely only within the very central regions of the cluster \citep{DML}.
Studies which stack many clusters to approximate a spherically symmetric
supercluster circumvent this difficulty, since average density becomes a
monotonic function of radius within $R_v$ \citep[e.g. ][]{B+97,B+98}.  In our case,
the outer regions of the clusters often contain several large groups or
other clusters of galaxies (see Fig.~\ref{fig-spatial}).  
Thus, it is probably more appropriate to consider
the local density of the galaxies as the most physically interesting variable.  

To compute the local density of cluster
members, we consider all galaxies in the spectroscopic catalogue (including those
with bad ADC or H$\alpha$ measurements) brighter than $M_b=-19$, and within
3$\sigma$ of the cluster redshift.
We then take the distance to the tenth nearest galaxy, in projected radius,
as $r_{10}$; the local projected density is then $\Sigma=10/\pi r_{10}^2$.
For galaxies near the boundary of a cluster catalogue, this will underestimate
the true density.  To partially account for this, we only consider galaxies within
18 Mpc of the cluster centre, so they are at least 2 Mpc from the edge of the
catalogue.  In some cases, however, the current \2df\ database is incomplete within
the 20 Mpc extracted area, and the densities of galaxies near these incomplete
regions will still be underestimated.

\begin{center}
\begin{figure}
\leavevmode \epsfysize=8cm \epsfbox{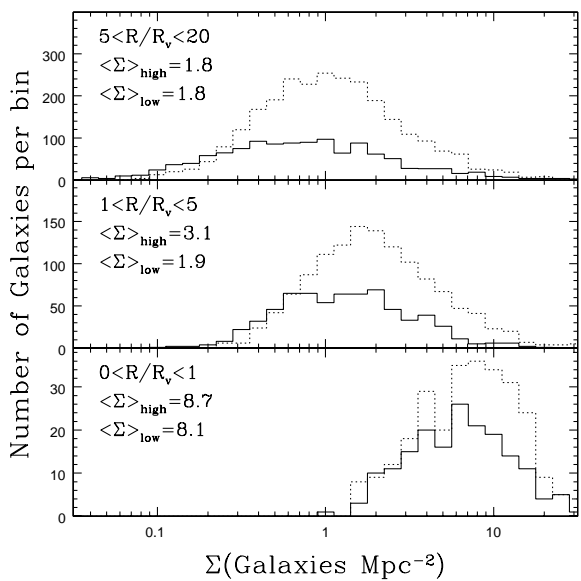}
\caption{The local, projected density distribution of galaxies in different
radial bins as labelled.  Clusters with $\sigma>800$ km s$^{-1}$ are shown as
the {\it dotted line}, and the mean value is shown in the top left corner 
as $\Sigma_{\rm high}$.  The {\it solid line} represents clusters with 
$\sigma<800$ km s$^{-1}$; their mean value is $\Sigma_{\rm low}$.
\label{fig-dendist}}
\end{figure}
\end{center}

In Fig.~\ref{fig-dendist} we show the distribution of density, for galaxies
in three radial bins.  In the cluster centre, almost all galaxies are in regions 
of very high local density.  However, at large radii galaxies can be found in a wide
range of environments; in particular it is not uncommon to find galaxies at $R>3R_v$ with
local densities as large as those within the virialized region.  Within the virial radius,
the distribution of $\Sigma$ is similar for both high and low velocity dispersion clusters;
the means are the same within $\sim 5$\%, and the probability that both distributions are
drawn from the same population is 0.12 as determined by a Kolmogorov-Smirnov test.
Between $1<R_v<5$, however, there is a significant ($>99.999$\%) difference, and
the mean local density of the clusters with $\sigma>800$ km s$^{-1}$ is more than twice as large as 
that of the lower velocity dispersion clusters.  

\begin{figure*}
\leavevmode \epsfysize=10cm \epsfbox{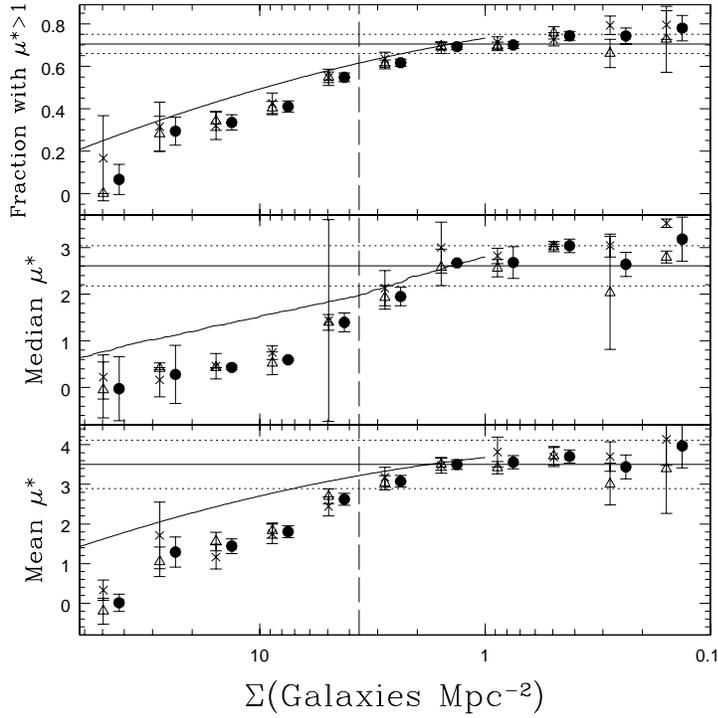}
\caption{As Figure~\ref{fig-sfrrad}, but as a function of local projected
density.  The vertical, {\it dashed line} represents the mean projected density of
galaxies within the virial radius of the cluster.  The solid curves (discussed in Section~\ref{sec-morph}) are the
expected trends due to the morphology-density relation of Dressler (1980), assuming
the field population is composed of 18\% E, 23\% S0 and 59\% spiral galaxies
\citep{WGJ}.
\label{fig-sfrden}}
\end{figure*}

In Fig.~\ref{fig-sfrden}, we show the properties of the cluster $\mu^\ast$ distribution,
as in Fig.~\ref{fig-sfrrad}, but plotted against $\Sigma$.  
The vertical line shows the mean projected density of galaxies within the virial radius,
$N(<R_v)/\pi R_v^2$ (note that this is not the same as the average of the $\Sigma$ values
calculated for each galaxy).
As in Fig.~\ref{fig-sfrrad}, the horizontal lines show the values of each statistic in the field.  
The field actually spans a range of densities, likely similar to that seen far
($>5 R_v$) from the cluster centre (see Fig.~\ref{fig-dendist}); however, our density
estimate in clusters is a measure of the galaxy density projected along a line-of-sight column of
unknown length, and thus cannot be directly applied to field galaxies to obtain
a comparable measurement of local density.
Thus, it is evident that star formation is suppressed at densities of $\Sigma\sim 1.5$
galaxies Mpc$^{-2}$, approximately 2.5 times
lower than the mean projected density of the cluster virialized region.

\begin{figure*}
\leavevmode \epsfysize=10cm \epsfbox{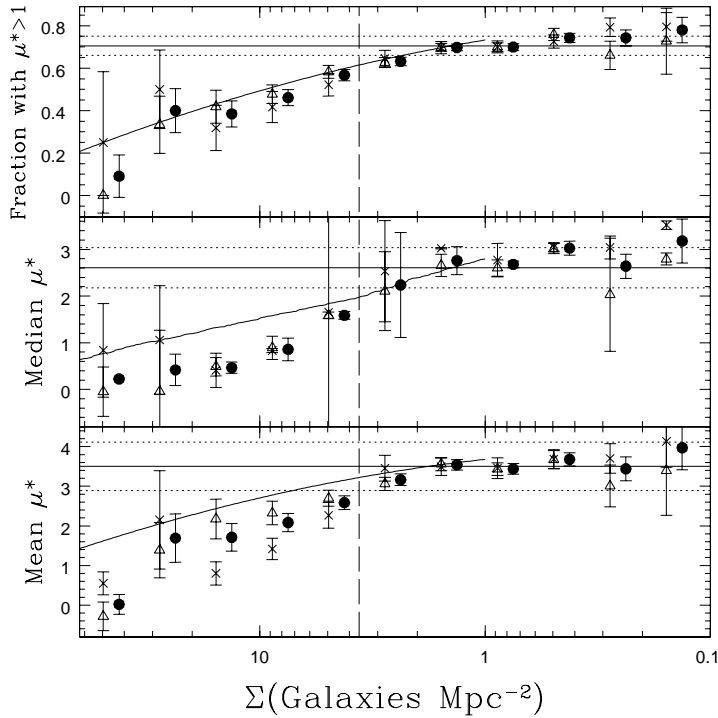}
\caption{As Figure~\ref{fig-sfrden}, but restricted to galaxies beyond $2R_v$.
\label{fig-sfrden2r}}
\end{figure*}

As in Fig.~\ref{fig-sfrrad}, the star formation rate distribution
at a given density is similar in both high- and low-velocity dispersion clusters.  This suggests that galaxy star
formation rates depend only on the local density, regardless of the larger-scale structure
in which they are embedded, although we repeat our caution that the velocity dispersions may
not be directly related to the cluster mass in all cases.
Furthermore, as we show in Fig.~\ref{fig-sfrden2r}, the correlation of star formation rate with 
density holds at $r>2R_v$, well outside the virialised cluster region.   This demonstrates that star formation
is low relative to the global average in {\it any} region exceeding the critical density
of 1 galaxy (brighter than $M_b=-19$) per Mpc$^{2}$, regardless of its proximity to 
a rich cluster \citep[see also][]{PG84}.  

\section{Discussion}\label{sec-discuss}
\subsection{Comparison with the morphology-density relation}\label{sec-morph}
We have shown that the dependence of star formation rate on local galaxy density
is independent of cluster velocity dispersion and thus, presumably, mass 
(see Section~\ref{sec-density}).  In a recent photometric
study based on {\it Hubble Space Telescope} imaging of 17 clusters, \citet{lowlx-morph}
found some evidence that the morphology-density relation {\it does} depend on cluster
X-ray luminosity, which is likely to be a better tracer of mass than velocity dispersion; 
at a given local density, low-mass clusters have more disk-dominated galaxies
than high-mass clusters.  Furthermore, they showed that this is most likely due to a difference
in the population of galaxy bulges; the disk luminosity function at a fixed
local density does not depend on cluster mass.  Since star formation is generally
limited to the galaxy disk, our results are consistent with this picture.  The luminosity
of a disk, and its star-forming activity, depend only on galaxy density, while the
luminosity of the bulge component has an additional, small dependence on the mass of the
embedding structure.

It would be of great interest to compare the dependence of $\mu^\ast$ on density with
the similar density-dependence of morphology, to determine the degree to which the two 
correlations are independent.  In particular, any difference between the two shows that
cluster galaxies differ from their morphological counterparts in the field, which supports the
hypothesis that they have undergone a physical transformation \citep{B+98}.
However, we note that this test is not conclusive; if the star formation rate of a spiral
galaxy is reduced gradually, on timescales similar to that for morphological change, 
the correlation between morphology and star formation rate may be retained, despite
the transformation.

Unfortunately, morphological classifications are not yet available for our sample.
However, we can use the local morphology-density relation computed by \citet{Dressler},
assuming that it is universal.  The luminosity limit of our sample ($M_b=-19$) is similar to
that of Dressler, $M_b\approx -19.2$, after accounting for the difference in cosmology and 
making the transformation  $M_b = M_V + 0.72(B-V)$, assuming an average galaxy colour $B-V=0.8$
\citep{F+95}.  Thus, 
our density measurements should be comparable.
We will assume that the field galaxy sample is composed of 18\% E, 23\% S0 and 59\% 
spiral and irregular galaxies \citep{WGJ,D+97}.  
We therefore divide the field galaxy $\mu^\ast$ distribution (Fig~\ref{fig-sfrdist}) into
three populations, identifying the lowest 18\% of $\mu^\ast$ values with the E population,
the next 23\% with the S0s, and the remainder with spirals.  It is then straightforward to
recompute the statistics shown in Fig.~\ref{fig-sfrden} for any morphological mix.
We show the expected $\mu^\ast$-density relation computed in this way, assuming Dressler's
morphology-density relation, as the solid curves in Fig.~\ref{fig-sfrden}.  Two things are
immediately clear.  First, at $\Sigma=1$ Mpc$^{-2}$, the lowest density point in Dressler's
study, the cluster morphological mix is close to that adopted for the field, so the predicted
curve is in good agreement with our measurements.  Note that this is dependent on an accurate
determination of the early-type fraction in the field, estimates of which have increased from
the 20\% adopted by \citet{Dressler}, to 30\% \citep{ST81}, adopted by
\citet{PG84}, and finally to the 41\% used here and elsewhere \citep{WGJ,D+97}.
This high value for the early-type fraction is a consequence of the bright luminosity limit,
and is consistent with that derived from type-dependent luminosity functions of \citet[][40\% at $M^\ast$]{marzke_cfa_morph}.
The second point is that the predicted $\mu^\ast$-density correlation appears to be shallower than
the observed relation.  This suggests that the morphology-density relation may be
distinct from the star formation-density relation.  In making this
comparison we have made the extreme assumption that the lowest values of $\mu^\ast$ are associated
with elliptical galaxies, and the highest values with spiral galaxies.  Any dispersion in the
natural morphology-$\mu^\ast$ relation will serve to further flatten the predicted $\mu^\ast$-density
relation and increase the discrepancy with the data.  On the other hand, 
there is an important caveat,
as \citet{Dressler} did not subdivide the late-type morphology class,
and Sa galaxies are known to have much less current star formation than irregular galaxies
\citep{K92,Jansen}.  If the fraction of Sa galaxies relative to later types increases with
density, this will steepen the curves in Fig.~\ref{fig-sfrden}.  

\subsection{Possible mechanisms: comparison with theoretical models}
These results show conclusively that suppressed star formation is not limited
to the cores of rich clusters, but is found in any environment in which the local
projected galaxy density exceeds one galaxy brighter than $M_b=-19$ per Mpc$^{-2}$.
This is in approximate agreement with the results of \cite{Kodama_cl0939}, though a
direct comparison is not possible because that survey probes much deeper down the
luminosity function, so the local projected galaxy densities are higher in the
same environments.  Whatever mechanism is responsible
for terminating star formation in galaxies, then, is not particular to the
cores of rich clusters, but is associated with dense groups in the cluster
infall regions as well.  This means that ram pressure stripping of galaxy disks
cannot be completely responsible for the correlation of star formation with
local density, since this is only expected to take place in the cores of
rich clusters \citep{GG,Fujita-rps,QMB}.

Most hierarchical models of galaxy formation do not include a calculation
of ram-pressure stripping of the cold, disk gas, nor of other physical processes like galaxy
harassment \citep{harass} which might play a role in dense environments.  The only
environmental effect on star formation rate in these models -- apart from a possible difference
in merging history -- is related to the hot, halo gas hypothesised to surround
every isolated galaxy.  It is assumed that galaxies maintain the
supply of cold gas -- fuel for star formation -- via continuous cooling from
a hot, diffuse gas halo associated with the dark matter potential \citep{SP99,KCDW,Cole2000}.  In haloes
with more than one galaxy, this hot gas is only associated with the central
galaxy; satellite galaxies are assumed to lose their supply of fresh fuel through
ram pressure stripping and tidal effects (though these are not directly modelled).  
In these models, therefore, star formation
rates begin to decline for any satellite galaxy, whether in a poor group or a
rich cluster.

\begin{figure}
\leavevmode \epsfysize=8cm \epsfbox{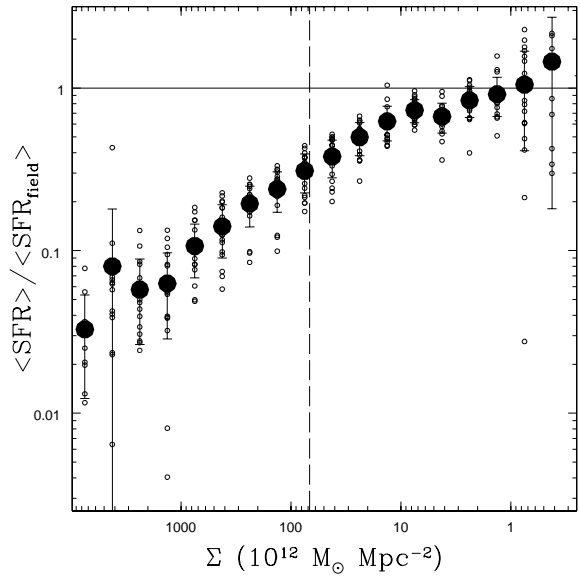}
\caption{The ``group'' model of \citet{infall}, in which galaxies in groups
with circular velocity $V_c>600$~km~s$^{-1}$ have their hot gas haloes stripped, so no further
cooling is permitted.  The density is the local projected mass density, 
computed from the area enclosing the nearest ten particles in projection.
Small, {\it open} points are results from a single projection of each of
six model clusters; the large {\it solid} point is the mean, and the error
bar is the standard deviation of the 18 realizations of the model.
The {\it dashed line} shows the mean density of particles within the virial
radius.
\label{fig-model}}
\end{figure}

These models are able to reproduce radial gradients in star 
formation within the virial radius of clusters to a remarkably high degree of
accuracy \citep{Diaferio,Okamoto}.
In particular, \citet{Diaferio} predict that the mean star formation rate should be
equivalent to the field value beyond $\sim 2 R_v$, in physical (i.e. not projected)
space.  
The model of \citet{infall} is a greatly simplified version of this more
complete model, as the properties of the field galaxy population are not modelled directly, but
are taken empirically from observations of the $z\sim 0.3$ field.  The advantage is that the effects of the
halo-stripping can be seen directly, since that is the only physical process (apart from
gravity) which is accounted for.   
In Fig.~\ref{fig-model}, we show the predictions of this
model, for the mean star formation rate relative to the
field, as a function of local projected density.  
The simulations on which the model is based were kindly provided
by Julio Navarro.  Here, local density is defined as the
projected surface mass density, computed by finding the
radius encompassing the ten nearest (in projection) particles in the simulations.  The model
is the ``group'' model in Fig. 1 of \citet{infall}; galaxies are assumed
to lose their reservoir of hot gas when they are associated with a group with
circular velocity $V_c>600~$km~s$^{-1}$.  While a direct comparison with the data is not
possible, since these simulations only provide the dark matter density, a comparison
relative to the mean surface density within $R_v$ should be fair if mass traces light.
First we note that the approximately power-law dependence on local projected density has a similar
slope in the data and the model; the mean star formation rate decreases by a factor
of $\sim 3$ for every factor 10 increase in surface density.  Secondly, in the model
the correlation flattens out at surface densities $\sim 1/7$ that of the mean projected density within
$R_v$.  Although this threshold is a factor $\sim 2$ lower than seen in the data, given the
crudity of the model, we consider the agreement reassuring.  
Unfortunately, the simulations used in this model did not include a large enough volume to
probe beyond a few $R_v$.  Thus, the low density regions in the simulations are not drawn
from the same regions in space as the low density regions in the observations, most of
which are found well beyond $2R_v$.

Thus, models in which halo-stripping is the only direct environmental-influence on
the galaxy star formation rate provide a reasonably good match to the data.
This is especially remarkable given that the stripping is not even directly modelled;
it is simply assumed that every satellite galaxy has {\it no} reservoir of hot
gas, immediately after it merges with a larger halo.  Improvement in this respect
alone may well improve the models' success in the lower density regions, far from
the cluster core.

\subsection{Consequences on the evolution of the global star formation rate}
What mechanisms are responsible for driving the strong observed evolution of
the global star formation rate \citep[e.g.][]{L96}?  One possibility is that
the decline in star formation activity is related to physics internal to
individual galaxies --- for example, consumption of a limited gas supply, or a
time-dependent cooling rate --- regardless of their environment.  On the other
hand, some of the decline is likely to be tied to the hierarchical growth of
structure; as time progresses, more and more galaxies are locked up in clusters
where, perhaps, star formation is directly inhibited.  

According to extended Press-Schechter theory \citep{PSext,B+91} the fraction of mass in
haloes greater than $10^{14}M_\odot$, approximately the limit of our cluster sample,
is only 11\% at the present day, and negligible by $z=1$.  Thus it is not
immediately obvious that the lower star formation rates in these systems can have
any effect on the global average.  However, we have shown that lower star formation
rates are seen in environments with densities $\sim0.3$ times lower than the mean
cluster density, regardless of their proximity to the cluster. 
This density corresponds approximately to the density {\it at} the
virial radius; by definition, if mass traces light then any virialised structure 
will have a mean density which exceeds this threshold.  Since our density estimate
is based on the tenth nearest galaxy brighter than $M_b=-19$, we cannot be sure
how our results apply to systems with fewer than ten such galaxies.  A virialised
system with more than ten galaxies brighter than this limit is expected to have a
total gravitational mass $M\gtrsim10^{13} M_\odot$, assuming a total mass-to-light
ratio of 100 \citep[e.g.][]{G+02}.  In contrast with the more massive clusters, these haloes account for $\sim 35$\% of the
mass in the present day Universe, and contribute significantly to
the global average star formation rate.  Furthermore, at $z=1$ only about 10\% of the
mass was in such environments; the rapid growth to $z=0$ on these mass scales may
well be able to explain the rapid evolution in the global star formation rate.
The hypothesis that the growth
of structure is largely responsible for the observed decline in star formation
with cosmic time \citep[e.g. ][]{L96} therefore becomes much more attractive.

\section{Conclusions}\label{sec-conc}
We have presented a study of seventeen known galaxy clusters, using redshifts and
H$\alpha$ equivalent widths measured from \2df\ spectra.  We have used this
to trace the dependence of relative star formation rates as a function of
radius and local density.  We conclude the following:
\begin{itemize}
\item[1.] The distribution of star formation rates is correlated with both distance
from the cluster centre, and with local projected density.  The distribution
becomes equivalent to that of the global average for radii $\gtrsim 3 R_v$,
and local projected densities $\lesssim0.3$
times that of the mean cluster virialized region.  These results are in good
agreement with preliminary results from the Sloan Digital Sky Survey \citep{Sloan_sfr}.
\item[2.] The correlation between star formation rate and local projected density holds
for galaxies more than two virial radii from the cluster centre.  Thus, star formation rates
depend primarily on the local density, regardless of their proximity to a rich cluster.
\item[3.] This means that galaxy transformation is not primarily driven by processes
like ram pressure stripping, which only operate in the most extreme environments,
but by processes which are effective in lower density, group environments.
\item[4.] 
The dependence
of star formation rate on density is the same for clusters with $\sigma>800$ km s$^{-1}$  and
for clusters with $\sigma<800$ km s$^{-1}$, which implies that the star formation rate
is insensitive to the global, large-scale structure in which the galaxy is embedded.
\item[5.] The correlation between star formation and density that is predicted from the
morphology-density relation of \citet{Dressler} is less steep than observed.
This provides conditional support for the view that the correlations with density are due to 
physical transformation of galaxies in dense regions, and that 
morphological change occurs on a different timescale from changes to the current
star formation rate.  However, it may also be explained by a lower fraction of
early type spiral galaxies, relative to late types.
\end{itemize}
\section*{Acknowledgements}
We thank an anonymous referee for useful comments.
MLB acknowledges support from a PPARC rolling grant for extragalactic
astronomy at Durham.  R.D.P. and W.J.C. acknowledge funding from the
Australian Research Council.
We thank Julio Navarro for providing the numerical simulations, and the
Sloan Digital Sky Survey collaboration for sharing their results in advance
of publication.
We gratefully acknowledge the support of the staff of the Anglo-Australian
Observatory for their assistance supporting 2dF throughout the
survey, and of the Australian and UK time assignment committees for
their continued support for this project.
\bibliography{ms}

\end{document}